\documentclass[twocolumn,english,aps,prl,superscriptaddress,amsmath,amsfonts]{revtex4}
\usepackage[T1]{fontenc}
\usepackage[latin9]{inputenc}
\usepackage{graphicx}

\usepackage{babel}

\begin{document}

\title{Universal Correlations and Dynamic Disorder in a Nonlinear Periodic
1D System}

\author{Yaron Silberberg}

\email[]{yaron.silberberg@weizmann.ac.il}

\affiliation{Department of Physics of Complex Systems, Weizmann Institute of Science,
Rehovot 76100, Israel}

\author{Yoav Lahini}

\affiliation{Department of Physics of Complex Systems, Weizmann Institute of Science,
Rehovot 76100, Israel}

\author{Yaron Bromberg}

\affiliation{Department of Physics of Complex Systems, Weizmann Institute of Science,
Rehovot 76100, Israel}

\author{Eran Small}

\affiliation{Department of Physics of Complex Systems, Weizmann Institute of Science,
Rehovot 76100, Israel}

\author{Roberto Morandotti}

\affiliation{Institute National de la Recherche Scientifique,
Varennes, Quebec, Canada}

\date{November 30, 2008}
\begin{abstract}
When a periodic 1D system described by a tight-binding model is uniformly
initialized with equal amplitudes at all sites, yet with completely
random phases, it evolves into a thermal distribution with no spatial
correlations. However, when the system is nonlinear, correlations
are spontaneously formed. We find that for strong nonlinearities,
the intensity histograms approach a narrow Gaussian distributed around
their mean and phase correlations are formed between neighboring sites.
Sites tend to be out-of-phase for a positive nonlinearity and in-phase
for a negative one. The field correlations take a universal shape
independent of parameters. This nonlinear evolution produces an effectively
dynamically disordered potential which exhibits interesting diffusive
behavior.
\end{abstract}
\maketitle

The tight binding approximation is one of the simplest models that
predict band structure and ballistic motion in periodic systems
\cite{bloch}. Using a tight-binding model with disorder, P.W.
Anderson was able to predict and study the effect of localization
in disordered lattices \cite{anderson}. A nonlinear version of the
tight-binding model, better known as the Discrete Nonlinear
Schrödinger Equation (DNLSE) has been used to study nonlinear
evolution in periodic systems, initially in the context of
periodic molecular and mechanical systems \cite{scott}, and
extensively in recent years to describe nonlinear propagation in
optical waveguide lattices \cite{christodoulides,lederer}, as well
as matter-waves in light-induced lattices\cite{BEC}. In
particular, the DNLSE explains the formation of nonlinear
intrinsic localized modes, also known as discrete solitons
\cite{chandjoseph} or discrete breathers \cite{flach}. Disorder
and localization have been studied experimentally recently both in
matter-waves \cite{aspect} \cite{ingucio} and in photonic lattices
\cite{segev,lahini}. An issue of particular interest and some
debate is the interplay between localization and nonlinearity
\cite{shepelyansky,flach1}.

Here we report on a new phenomenon that results from the interplay
of nonlinearity and disorder. This effect occurs in periodic systems,
where nonlinearity induces disorder in an otherwise perfectly ordered
lattice. 
What we find is that when the system is initialized with random-phase
fields, it evolves into particular distributions with well defined
stationary statistical properties. Most interestingly, the field correlation
function and the distribution of phases assume universal forms independent
of the exact value of the nonlinear parameter. The resulting distribution
induces a dynamic structure with several intriguing properties.

%
\begin{figure}
\includegraphics[clip,width=0.9\columnwidth]{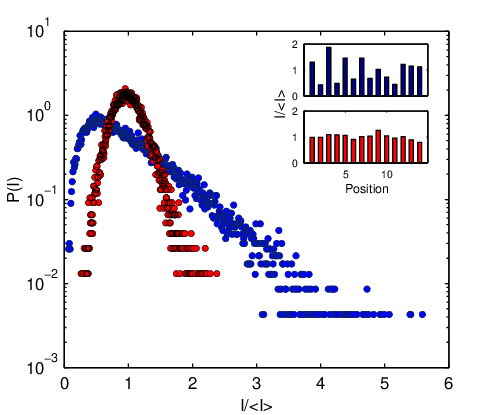}

\caption{Experimental measurement of output light intensities from a waveguide
array, when input fields with equal amplitudes, yet random phases,
are coupled to several adjacent waveguides. Measured values of many
random-phase realizations are shown as histograms for linear propagation
(blue circles) and high-intensity, nonlinear propagation (red circles).
The inset shows two sets of measurements for an input with the same
phase realization after linear (blue) and nonlinear (red) propagation.
The output intensity distribution is exponential-like in linear propagation,
and becomes narrow, Gaussian-like in the nonlinear case.}

\end{figure}

In Fig. 1 we show the results of an optical experiment in a waveguide
array that motivated this study. Light with uniform intensity yet
random phases was injected into a large number of waveguides in a
periodic waveguide lattice. The intensity at the output end was measured,
and the histograms of intensity values obtained from many repeats
of the experiment with different random phase realizations are shown
for both low intensity and high intensity light. For experimental
details, see \textit{Methods} below. While linear propagation produced
exponential-like distribution of intensities, as expected for summing
of many random-phase inputs, the nonlinear propagation yielded a much
narrower distribution around the average intensity. Increasing the
optical power led to the narrowing of the output distribution. Note
that for both measurements, the counts at low intensity values are
underestimated because of scattered light and instrumentation noise.
To investigate this behavior we have modeled the problem using the
DNLSE. While we use here the notations of optics, our results will
hold in general for all other systems described by the DNLSE. The
evolution of light in a periodic array of weakly coupled waveguides
is described by:

\begin{equation}
i\frac{da_{n}}{dz}=C(a_{n-1}+a_{n+1})+\gamma|a_{n}|^{2}a_{n},\label{DNLSE1:eq}\end{equation}
 where $a_{n}$ is the amplitude of the mode in the $n^{th}$ waveguide,
C is the coupling coefficient to the nearest neighbors and $\gamma$
is the nonlinear coefficient, positive (negative) for focusing (defocusing)
nonlinearity. We shall consider the situation where light is injected
into the array with uniform amplitudes $|a_{n}|=a_{0}$, yet with
completely uncorrelated, random phases. It is convenient to use the
normalized equation, \begin{equation}
i\frac{du_{n}}{d\zeta}=(u_{n-1}+u_{n+1})+\Gamma|u_{n}|^{2}u_{n},\label{DNLSE2:eq}\end{equation}
 where $\zeta=zC$, $u_{n}=a_{n}/a_{0}$ and $\Gamma=\gamma{a_{0}}^{2}/C$.
With this normalization, the input intensities are all uniform with
$I_{n}(0)=u_{n}{u_{n}}^{*}=1$.

Consider first the linear problem, i.e. $\Gamma=0$. As might be expected,
after a certain distance, mixing of the different input fields leads
to an output pattern with fluctuating intensities. Fig. 2(a)-(c) shows
results of numerical simulations of Eq. (2) for various properties
of the fields after propagating a distance of $\zeta=10$ in an array
with $N=256$ waveguides. Periodic boundary conditions are used to
avoid edge effects. The results shown are averaged over 500 realizations
with different random initial fields. Fig. 2(a) shows the intensity
histogram; it follows an exponential law, $P(I)=\exp(-I)$, as expected
for random fields. Fig. 1(b) shows the field correlations $C_{k}=\sum(u_{n}u_{n+k}^{*}+u_{n}^{*}u_{n+k})/2N$
demonstrating, as could be expected, that the fields at different
sites do not correlate. Finally, Fig 1(c) shows the histogram of phase
differences between neighbors, $\theta_{n}=\phi_{n}-\phi_{n+1}$,
with $\phi_{n}=\arg u_{n}$ the phase of the field $u_{n}$. These
phases are uniformly distributed.

%
\begin{figure}
\includegraphics[width=1\columnwidth]{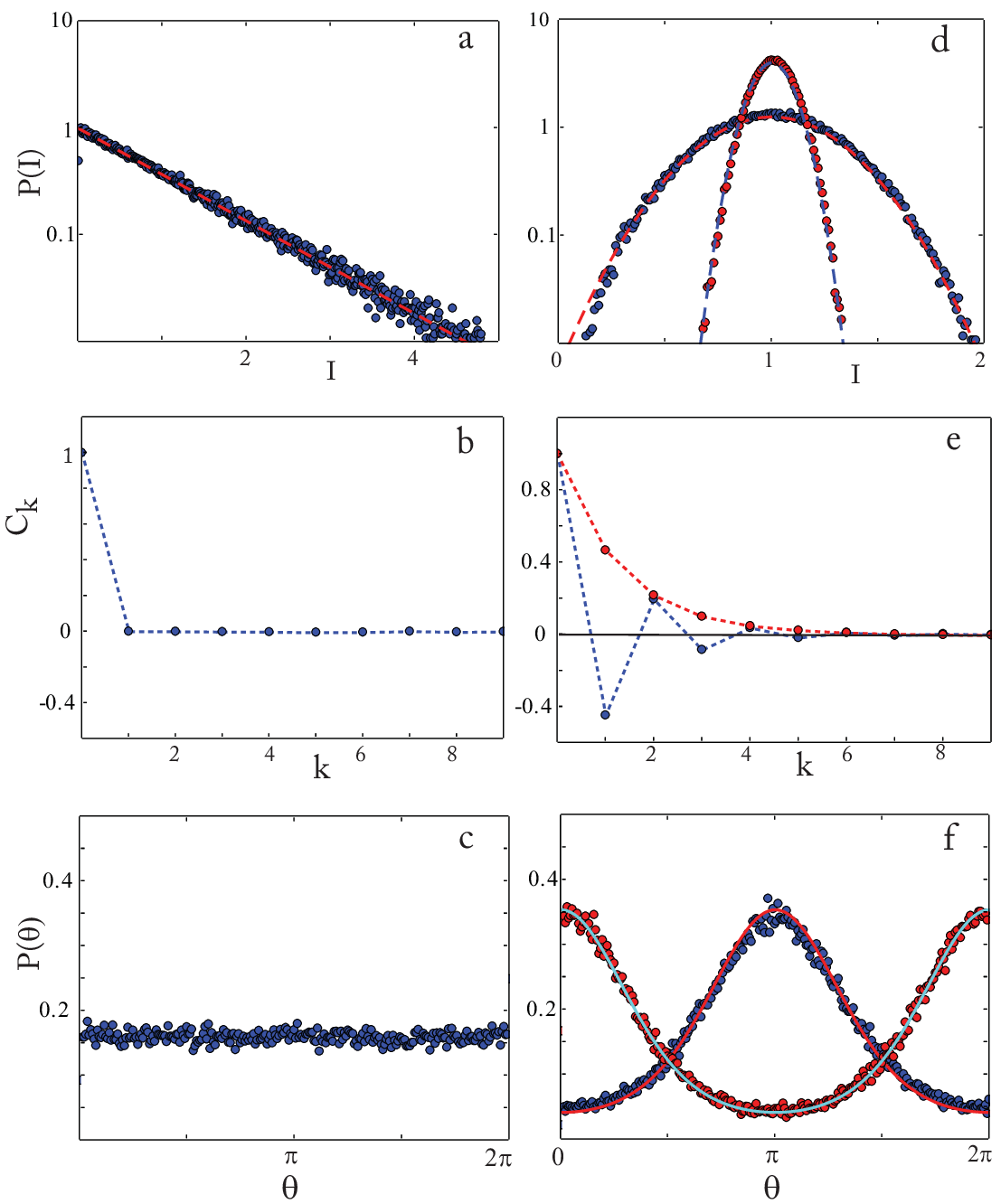}

\caption{Simulation results of the DNLSE with uniform intensities and random
phases. (a) Intensity histogram, (b) field correlation and (c) phase-difference
histogram for a linear system ($\Gamma=0$), exhibiting exponential
intensity distribution and uncorrelated fields and phases. (d) Intensity
histograms for $\Gamma=\pm20$ (blue circles) and $\Gamma=\pm200$
(red circles). The distributions are narrower at higher nonlinearities,
yet independent of the sign of $\Gamma$. (e) The universal field
correlation function is identical for both nonlinear values, but depends
on their sign: blue for positive (focusing, $\Gamma>0$), red for
negative nonlinearities. (f) Phase difference histograms also approach
a universal distribution, concentrating around $\pi$ for positive
nonlinearity and 0 for the negative case. The theoretical lines in
(d)-(f) are predictions of Eqs. (7) and (8).}

\end{figure}

We now repeat the simulations with nonlinearity, and we will be interested
mostly in the limit of strong nonlinearity; results for $\Gamma=\pm20$
and for $\Gamma=\pm200$ are given in Figs. 2(d)-(f). Two changes
from the linear case are obvious. First, the intensity histograms,
shown in Fig. 2(d), now converge around the average intensity value
of 1, with a width that shrinks with the nonlinear parameter. The
distribution seems to fit well a gaussian distribution with $P(I)=\exp[-(I-1)^{2}/2\sigma^{2}]$,
and it is independent of the sign of the nonlinearity.

The second major effect of the nonlinearity is the induced spatial
field correlations. Most interestingly, the correlation function (Fig.
2(e)) takes a shape that is independent of the nonlinearity value,
and is sensitive only to its sign. It is described well by exponential
decays, $C_{k}=(-1)^{k}\exp(-\alpha k)$ for positive (focusing) nonlinearity
and $C_{k}=\exp(-\alpha k)$ for the negative case. Note that the
correlation is only visible in the fields - the intensities remain
uncorrelated; Intensity correlations show a diminished peak at k=0
and a uniform background. These field patterns are consistent with
the known properties of modulation instability in such systems: Staggered
(unstaggered) fields are stable in positive (negative) nonlinearity
arrays \cite{modulation}.

Since the intensities become more uniform at high nonlinearities,
the field correlation functions are dictated by the variations of
phases between neighboring sites. In Fig. 1(f) we show the histograms
of these phase differences for positive and negative nonlinearities.
While in the former neighboring waveguides are most likely to be out
of phase, as the distribution peaks at $\pi$, in the latter two neighboring
waveguides tend to be in-phase. This distribution of phases also attains
a constant profile at high nonlinearity values.

The field correlation and the intensity distribution are closely related.
This can be deduced from the conserved quantities, the Hamiltonian
\begin{equation}
H=\frac{1}{2}\Gamma\sum I_{n}^{2}+\sum(u_{n}u_{n+1}^{*}+u_{n}^{*}u_{n+1}),\label{H}\end{equation}
 and the total photon number, \begin{equation}
A=\sum I_{n}.\label{A}\end{equation}

From these it is easy to show that \begin{equation}
\frac{\Gamma}{4}\sigma^{2}(\zeta)+C_{1}(\zeta)=H_{0}\label{cons}\end{equation}
 is also constant. Here $\sigma=(\Sigma I_{n}^{2}/N-1)^{1/2}$ is
the standard deviation of the intensities. Since our initial condition
are of uniform intensities, $\sigma^{2}(0)=0$, and random phases,
$C_{1}(0)\approx0$, then $H_{0}=0$, hence \begin{equation}
\frac{\Gamma}{4}\sigma^{2}(\zeta)=-C_{1}(\zeta).\label{dev}\end{equation}
 Equation (6) predicts that the signs of $C_{1}$ and $\Gamma$ are
different, as indeed is observed in Fig 2. For weak nonlinearity ($\Gamma\ll1$),
when the distribution deviate only slightly from exponential, a small
correlation is formed with $C_{1}=-\Gamma/4$. However, as the nonlinearity
increases, Eq. (\ref{dev}) predicts that the intensity distribution
has to narrow down, since necessarily $|C_{1}|<1$. 

To relate the phase and intensity fluctuations, we have to study the
statistical properties of the DNLSE. Such investigations were carried
out by Rasmussen et al \cite{rasmussen}, and extended later by Rumpf
\cite{rumpf}. They were mostly interested in the conditions for the
generation of localized structures. With our initial conditions, it
can be shown that localized structures are not formed, but we can
use the same formalism to derive the probability distributions for
phase and intensities.

In essence, the state of maximal entropy $S[p(I_{1},...I_{N},\theta_{1},...\theta_{N})]=-\int p\ln p\prod dI_{i}d\theta_{i}$
can be derived by the variational problem $\delta(S-\alpha A-\eta H-\lambda\int p)=0$,
where $\alpha,\eta$ and $\lambda$ are the appropriate Lagrange multipliers
\cite{rumpf}. The problem is greatly simplified by the approximation
$\Sigma(u_{n}u_{n+1}^{*}+c.c.)\approx2\sum\cos(\theta_{n})$, which
is consistent with the observation that the intensities are uncorrelated
and at high nonlinearities they are close to their average value of
1. With this approximation, the intensities and phases are separable,
and their distributions are derived to be: \begin{eqnarray}
p_{I}(I) & = & \lambda_{1}\exp[-\frac{\eta\Gamma}{2}(I-1)^{2}]\\
p_{\theta}(\theta) & = & \lambda_{2}\exp[-2\eta\cos(\theta)]\end{eqnarray}
 with $\lambda_{1},\lambda_{2}$ appropriate normalization constants
and $\eta\approx\pm0.533$ is the solution of $4\eta\int\cos(\theta)p_{\theta}(\theta)d\theta+1=0$,
where the sign is selected to match the sign of $\Gamma$. The phase
distribution is then maximized at $\theta=\pi$ ($\theta=0$) for
positive (negative) $\Gamma,$ respectively. Note that these universal
phase distribution functions, which are shown as lines in Fig 2(f),
are independent of the value of the nonlinearity. They lead to the
universal correlation function that decays with $C_{k}/C_{k+1}=-4\eta$
as shown in Fig 2(e).

Finally, we would like to discuss the potential (or index) pattern
which is induced by the fluctuating fields through the nonlinear effect.
The fields induce an effective dynamic disorder into the lattice;
While we have shown above that the variations in intensities tend
to diminish at high nonlinearities {[}$\sigma=(\eta\Gamma)^{-1/2}$],
the variations in induced nonlinear potential, i.e. $\Gamma\sigma$
actually increase. The array becomes effectively more disordered:
A stationary structure with this level of disorder would have been
characterized by a localization length that is narrower than the lattice
spacing. However, this increase in disorder is also accompanied by
faster dynamics: The induced pattern changes faster at higher nonlinearities.
The nonlinear potential map $\Gamma I_{n}(\zeta)$ is shown in Fig.
3, together with the intensity correlation maps $Y(k,\delta\zeta)=\Sigma_{n}\int d\zeta I_{n}(\zeta)I_{n-k}(\zeta-\delta\zeta)$.
Note that while at a given $\zeta$ the intensities at adjacent waveguide
do not correlate, $Y(k,0)=\delta_{k,0}$, they are correlated at other
values of $\zeta$. It is this dynamic potential structure that determines
the field correlations and other peculiarities of this system. 
%
\begin{figure}
\includegraphics[width=1\columnwidth]{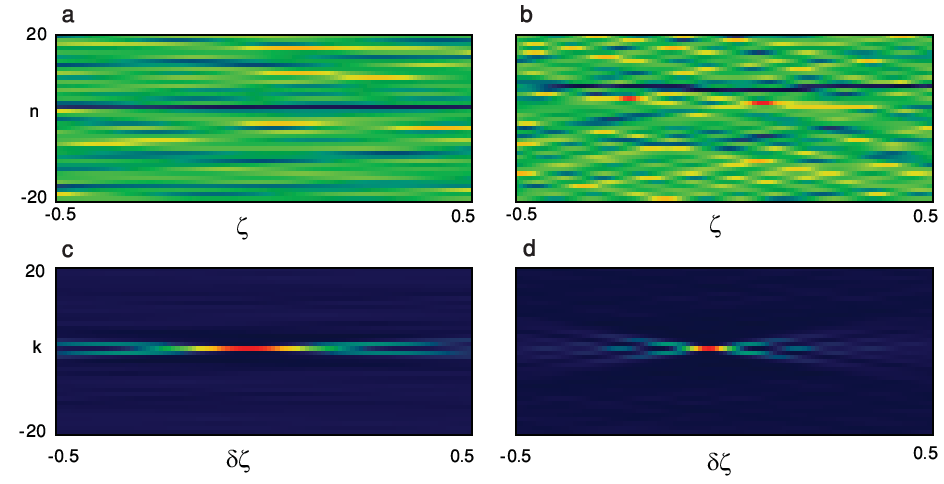}\\

\caption{Maps of nonlinear potential induced by the fluctuating fields for
(a) $\Gamma=20$ and (b) $\Gamma=200$ and the corresponding correlation
maps $Y(k,\delta\zeta)$ (c,d). The maps show a section of 40 waveguides
(vertically) and a propagation of $\zeta=1$. Note the faster dynamics
for the higher nonlinearity, evident also in the correlation map.
For easy viewing, the potentials are normalized to their peaks, although
the values in (b) are about 3 times higher than in (a)}

\end{figure}

One way to characterize this structure is to investigate the transport
properties of a a secondary probe beam that propagates in the nonlinearly
induced disordered potential. We have simulated the propagation of
such a weak probe field in an array with 1024 sites. The probe is
launched into a single central site at $\zeta=0$, and its width is
monitored along propagation. Fig. 4(a) shows this width, averaged
over several realizations, as a function of $\zeta^{1/2}$, for two
values of nonlinearity, $\Gamma=20$ and $\Gamma=200$, and Fig 4(b)
shows the probe intensity distribution at $\zeta=2000$. It seems
that the broadening is governed mostly by diffusive dynamics, that
is, a gaussian-like distribution that broadens diffusively as $\zeta^{1/2}$.
Indeed, dynamic disorder that is spatially uncorrelated is known to
lead to diffusive broadening \cite{dynamic1,dynamic2}. What we find
interesting is that at higher nonlinearities the diffusion coefficient
is actually larger, in spite of the stronger disorder. This is most
likely the result of the faster dynamics, but it could be that the
nontrivial correlation maps also plays a role. These points are currently
under investigation. 
\begin{figure}
\includegraphics[clip,width=0.75\columnwidth]{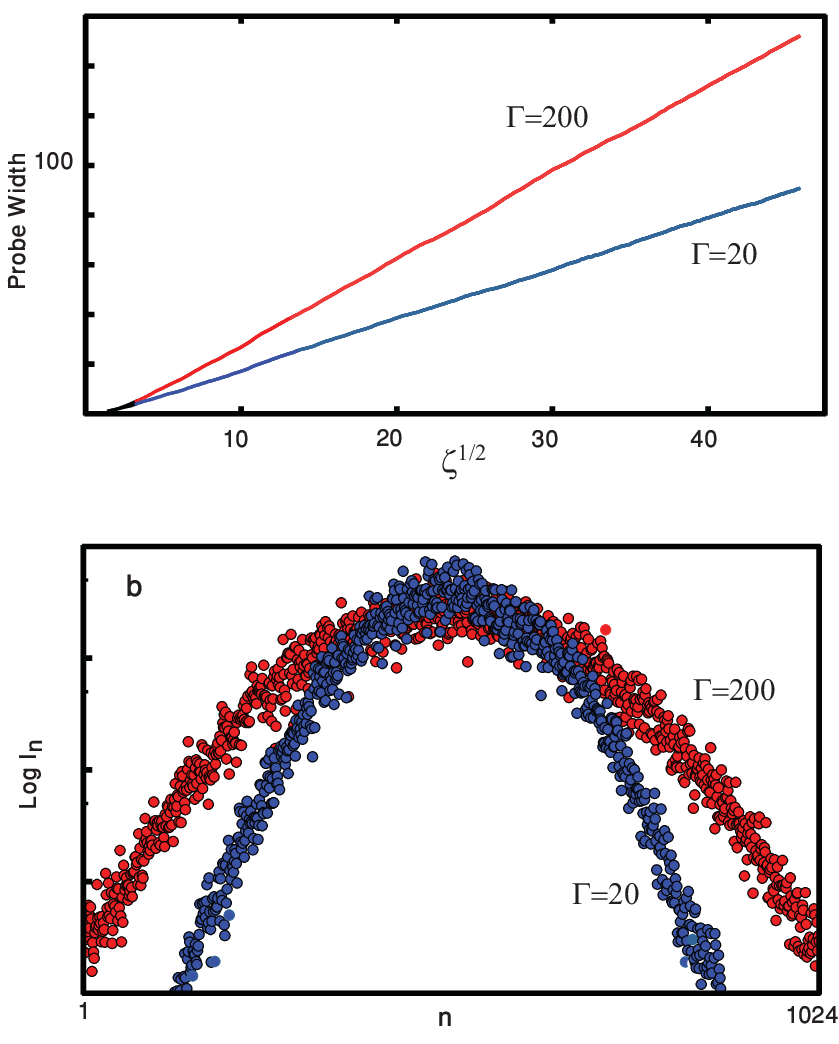}\\

\caption{The diffusion of a weak probe in the nonlinear potential induced by
the fluctuating fields with $\Gamma=20$ and $\Gamma=200$. (a) the
averaged probe width as a function of $\zeta^{1/2}$. (b) The averaged
probe profile at $\zeta=2000$}

.
\end{figure}

In conclusion, we have shown that when systems described by the DNLSE
are initialized with equal amplitudes yet random phases, universal
field and phase correlation are formed in the high nonlinearity limit.
In contrast with the thermal distribution of intensities obtained
in linear propagation, the intensity variations are diminished, and
universal phase correlations are formed. The underlying potential
field leading to this behavior is of a dynamic disordered system,
which is characterized by diffusive propagation.

\textit{Methods}: The periodic waveguide lattices were prepared on
an AlGaAs substrate using e-beam lithography, followed by reactive
ion etching. The width of each of the waveguides as well as their
spacing were 4 microns. The tunneling parameter between sites C was
measured to be $280m^{-1}$. The light source was a pulsed optical
parametric amplifier, producing 1.2 ps pulses at a wavelength of 1530nm
with a 1 KHz repetition rate. The spatial phase of the laser light
beam was randomized using a computer-controlled spatial light modulator.
The beam was shaped using a cylindrical optics and then injected into
the input facet of the array, covering ~15 lattice sites. The light
intensity profile was measured at the output facet of the 8mm long
sample using an infrared camera and the intensity statistics was calculated
by collecting many such images with random phase realizations.
\begin{acknowledgments}
We thank R. Helsten for valuable technical help. This work was supported
by the German-Israel Foundation (GIF) and the Crown Photonic Center.
YL is supported by an Adams Fellowship of the Israeli Academy of Science
and Humanities.
\end{acknowledgments}

\begin{thebibliography}{19}
\bibitem{bloch} F. Bloch, Zeits. Physik \textbf{52}, 555 (1928).

\bibitem{anderson} P. Anderson. 
Phys. Rev. 109, 1492 - 1505 (1958).

\bibitem{scott} J. C. Eilbeck, P. S. Lomdahl, and A. C. Scott, 
Physica D \textbf{16}, 318-338 (1985).

\bibitem{lederer} F. Lederer, G.I. Stegeman, D.N. Christodoulides,G.
Assanto, M. Segev, Y. Silberberg, Phys. Rep. \textbf{463}, 1-126
(2008).

\bibitem{christodoulides} D. N. Christodoulides, F. Lederer, and
Y. Silberberg, Nature (London) \textbf{424}, 817 (2003).

\bibitem{BEC} A. Trombettoni and A. Smerzi, Phys. Rev. Lett. \textbf{86},
2353 (2001).

\bibitem{chandjoseph} D. N. Christodoulides and R. I. Joseph, Opt.
Lett. \textbf{13}, 794 (1988).

\bibitem{flach} S. Flach and C.R. Willis, Phys. Rep. \textbf{295}, 181(1998).

\bibitem{aspect} J. Billy, V. Josse, Z. Zuo, A. Bernard, B. Hambrecht,
P. Lugan, D. Clement, L. Sanchez-Palencia, P. Bouyer, and A. Aspect
Nature \textbf{453}, 891-894 (2008).

\bibitem{ingucio} G. Roati, C. D'Errico, L. Fallani, M. Fattori,
C. Fort, M. Zaccanti, G. Modugno, M. Modugno, and M. Inguscio. 
Nature \textbf{453}, 95-898 (2008).

\bibitem{segev} T. Schwartz, G. Bartal, S. Fishman, and M. Segev.
Nature \textbf{446}, 52-55 (2007).

\bibitem{lahini} Y. Lahini, A. Avidan, F. Pozzi, M. Sorel, R. Morandotti,
D.N. Christodoulides, and Y. Silberberg, Phys. Rev. Lett.
\textbf{100}, 013906 (2008).

\bibitem{shepelyansky} A.S. Pikovsky and D.L. Shepelyansky, Phys. Rev. Lett.
\textbf{100}, 094101 (2008).

\bibitem{flach1}G. Kopidakis, S. Komineas, S. Flach, S. Aubry, Phys.
Rev. Lett. \textbf{100}, 084103(2008)

\bibitem{modulation} J. Meier, G. I. Stegeman, D. N. Christodoulides,
Y. Silberberg, R. Morandotti, H. Yang, G. Salamo, M. Sorel, J. S.
Aitchison, 
Phys. Rev. Lett. \textbf{92}, 163902 (2004).

\bibitem{rasmussen} K.O. Rasmussen, T. Cretegny, P.G. Kevrekidis.
and Gronbech-Jensen, N.,
Phys. Rev. Lett. \textbf{84}, 3740 (2000).

\bibitem{rumpf} B. Rumpf. 
Phys. Rev. E \textbf{77}, 036606 (2008).

\bibitem{dynamic1} A. Madhukar and W. Post, Phys. Rev. Lett. \textbf{39},
1424 (1977).

\bibitem{dynamic2} N. Lebedev, P. Maass and S. Feng, Phys. Rev. Lett.
\textbf{74}, 1895 (1995).
\end{thebibliography}

\end{document}